\documentclass[aps,twocolumn,nofootinbib,superscriptaddress,floatfix]{revtex4}

\usepackage{color,amsmath,amssymb,graphicx,latexsym,subfigure}
\usepackage{float}
\usepackage{booktabs}
\usepackage{threeparttable}
\usepackage[colorlinks,linkcolor=green, anchorcolor=green,citecolor=green]{hyperref}
\usepackage{ulem,soul}
\def\d{\mathrm{d}}

\newcommand{\be}{\begin{equation}}
\newcommand{\ee}{\end{equation}}
\newcommand{\ba}{\begin{eqnarray}}
\newcommand{\ea}{\end{eqnarray}}

\newcommand{\gsim}{\mathrel{\hbox{\rlap{\lower.55ex \hbox {$\sim$}}
			\kern-.3em \raise.4ex \hbox{$>$}}}}
\newcommand{\lsim}{\mathrel{\hbox{\rlap{\lower.55ex \hbox {$\sim$}}
			\kern-.3em \raise.4ex \hbox{$<$}}}}

\providecommand{\kwd}[1]
{
  \small	
  \textbf{\textbf{Keywords:}} #1
}

\hypersetup{colorlinks=true,
	breaklinks=true,
	pdfstartview=Fit,
	linkcolor=blue,
	citecolor=blue,
	urlcolor=blue}

\begin{document}

\title{Limits on scalar-induced gravitational waves from the stochastic background by pulsar timing array observations}

\author{Yi-Fu Cai} \email{yifucai@ustc.edu.cn}
\affiliation{Deep Space Exploration Laboratory/Department of Astronomy, School of Physical Sciences, University of Science and Technology of China, Hefei 230026, China}
\affiliation{School of Astronomy and Space Science, University of Science and Technology of China, Hefei 230026, China}

\author{Xin-Chen He} 
\affiliation{Deep Space Exploration Laboratory/Department of Astronomy, School of Physical Sciences, University of Science and Technology of China, Hefei 230026, China}
\affiliation{School of Astronomy and Space Science, University of Science and Technology of China, Hefei 230026, China}

\author{Xiao-Han Ma} 
\affiliation{Deep Space Exploration Laboratory/Department of Astronomy, School of Physical Sciences, University of Science and Technology of China, Hefei 230026, China}
\affiliation{School of Astronomy and Space Science, University of Science and Technology of China, Hefei 230026, China}

\author{Sheng-Feng Yan}
\affiliation{Deep Space Exploration Laboratory/Department of Astronomy, School of Physical Sciences, University of Science and Technology of China, Hefei 230026, China}
\affiliation{School of Astronomy and Space Science, University of Science and Technology of China, Hefei 230026, China}
\affiliation{Istituto Nazionale di Fisica Nucleare (INFN), Sezione di Milano, Milano 20146, Italy}
\affiliation{DiSAT, Universit\`{a} degli Studi dell'Insubria, Como 22100, Italy}
\affiliation{School of Physics, The University of Electronic Science and Technology of China, Chengdu 611731, China}

\author{Guan-Wen Yuan} 
\affiliation{School of Astronomy and Space Science, University of Science and Technology of China, Hefei 230026, China}
\affiliation{Key Laboratory of Dark Matter and Space Astronomy, Purple Mountain Observatory, Chinese Academy of Sciences, Nanjing 210023, China}
\affiliation{Institute of High Energy Physics, Chinese Academy of Sciences, Beijing 100049, China}

\begin{abstract}

Recently, the NANOGrav, PPTA, EPTA, and CPTA collaborations independently reported their evidence of the Stochastic Gravitational Waves Background (SGWB). While the inferred gravitational-wave background amplitude and spectrum are consistent with astrophysical expectations for a signal from the population of supermassive black-hole binaries (SMBHBs), the search for new physics remains plausible in this observational window. In this work, we explore the possibility of explaining such a signal by the scalar-induced gravitational waves (IGWs) in the very early universe. We use a parameterized broken power-law function as a general description of the energy spectrum of the SGWB and fit it to the new results of NANOGrav, PPTA and EPTA. We find that this approach can put constraints on the parameters of IGW energy spectrum and further yield restrictions on various inflation models that may produce primordial black holes (PBHs) in the early universe, which is also expected to be examined by the forthcoming space-based GW experiments. \\
\par\kwd{pulsar timing array observation, Bayesian inference, stochastic gravitational wave background, induced gravitational waves, early universe}
\end{abstract}

\date{\today}

\maketitle

\section{Introduction}

Gravitational waves (GWs), predicted by General Relativity and observed in binary pulsar systems~\cite{Detweiler:1979wn} as well as stellar-mass black hole merger events~\cite{LIGOScientific:2016aoc}, have provided a key tool for us to explore the very early universe, marking the dawn of multi-messenger astronomy.  
The bare interaction between GWs and primordial matter fields allows for the capture of pristine information from the primordial universe, uncontaminated by other astrophysical processes. 
In contrast, the nonlinearity of gravity and various astrophysical phenomena tend to obscure much of the primordial information carried by electromagnetic waves throughout the evolution of the universe. Fortunately, the photons from the cosmic microwave background (CMB) radiation can provide valuable insights into the origins of the universe. 

As a prevailing paradigm of the very early universe, inflationary cosmology~\cite{Starobinsky:1980te,Guth:1980zm,Linde:1981mu,Starobinsky:1982ee} provides a natural and compelling theoretical explanation for the anisotropies observed in CMB. During inflation, primordial density fluctuations were produced quantum mechanically and then were stretched to cosmological scales due to the exponential expansion of spacetime. These fluctuations later evolve to form a nearly scale-invariant, quasi-adiabatic, and nearly Gaussian power spectrum on large scales with $k < 1~\mathrm{Mpc}^{-1}$. 
The statistical properties of these perturbations can offer insights into the early stages of inflation, covering approximately 7 e-folds of its beginning. However, $\sim 55$ e-folds of the inflation phase are required to account for the universe we see today, leaving $\mathcal{O}(50)$ e-folds of inflation still uncertain. 
Fluctuations that exit the Hubble radius during the later stages of inflation correspond to the small-scale anisotropies in our universe, which have been dominated by nonlinear astrophysical processes. On the other hand, GW signals produced before the hot Big Bang, in principle, can provide information about phenomenology on small scales. 
According to the cosmological perturbation theory (see Ref.~\cite{Kodama:1984ziu, Mukhanov:1990me, Lyth:1998xn} for a review), any sufficiently violent process occurring in the early universe would contribute to GW signals that echo throughout the whole universe. 
The superposition of these GW signals, arising from different physical origins and produced during different periods is considered one of the most intriguing targets in current and future searches for GWs. 

Recently, collaborations of Pulsar Timing Arrays (PTAs), including NANOGrav~\cite{NANOGrav:2023gor}, Parkers PTA (PPTA)~\cite{Reardon:2023gzh}, European PTA (EPTA)~\cite{Antoniadis:2023ott} and the China PTA (CPTA)~\cite{Xu:2023wog}, have collectively announced compelling evidence supporting the existence of a stochastic gravitational wave background (SGWB) within the frequency range of approximately $1 - 10 ~\mathrm{nHz}$, which attracts lots of implications (e.g the reviews~\cite{NANOGrav:2023hvm, Antoniadis:2023xlr}). 
The previous release of the NANOGrav 12.5 years dataset presented a subtle trace of the SGWB, sparking scientific interest and prompting discussions on the potential origin of this signal within the nano-Hertz frequency range. The inspiraling of supermassive black hole binaries (SMBHBs) has been extensively discussed as one of the primary astrophysical candidate sources for generating the SGWB~\cite{Burke-Spolaor:2018bvk, Sesana:2008mz,Shen:2023pan,Broadhurst:2023tus} in the nano-Hertz frequency band. Numerous cosmological phenomena, beyond the inspiral of SMBHBs, can serve as potential sources of SGWB within the specified frequency band~\footnote{NOTE: many phenomenological papers appeared immediately after the announcement of the collaborations~\cite{Shen:2023pan, Franciolini:2023wjm, Lambiase:2023pxd, Han:2023olf, Guo:2023hyp, Wang:2023len, Ellis:2023tsl, Vagnozzi:2023lwo, Fujikura:2023lkn, Kitajima:2023cek, Franciolini:2023pbf, Megias:2023kiy, Ellis:2023dgf, Bai:2023cqj, Yang:2023aak, Ghoshal:2023fhh, Deng:2023btv, Mitridate:2023oar, King:2023cgv, Zu:2023olm, Li:2023yaj, Addazi:2023jvg, Liu:2023ymk, Konoplya:2023fmh, Unal:2023srk,Broadhurst:2023tus,Bian:2023dnv,Madge:2023cak,Balaji:2023ehk}}. 
For example, enhanced scalar perturbations at small scales during inflation might lead to the formation of primordial black holes (PBHs), which are potential dark matter candidates~\cite{Ivanov:1994pa, Carr:2016drx, Carr:2020xqk, Chen:2019xse, Yuan:2023bvh}. Simultaneously, these scalar perturbations can give rise to scalar-induced gravitational waves (IGWs)~\cite{Boyle:2005se, Ananda:2006af,Baumann:2007zm, Yuan:2021qgz,Saito:2009jt,Wang:2019kaf,Zhao:2022kvz, Papanikolaou:2020qtd} (see Ref.~\cite{Domenech:2021ztg} for a recent review). Additionally, cosmological first-order phase transitions and the topological defects such as cosmic strings and domain walls could also generate SGWB~\cite{Kosowsky:1992rz, Kamionkowski:1993fg,Caprini:2007xq,Hindmarsh:2013xza,Kibble:1976sj,Vilenkin:1981bx,Hogan:1984is,Caldwell:1991jj,Vilenkin:1981zs,Chang:1998tb,Hiramatsu:2010yz,Chen:2022azo,Ashoorioon:2022raz,Bian:2022qbh,He:2023ado}.
The SGWB generated by cosmological sources holds significant potential for advancing our understanding of new physics beyond the Standard Model and providing insights into the primordial universe. 

In this study, we focus on verifying the possibility of an enhanced scalar perturbation power spectrum as the source of the SGWB via analyzing the latest released dataset of NANOGrav, PPTA, and EPTA. Our objective is to identify statistical estimates for the parameters that influence the shape of the energy spectrum $\Omega_{\text{GW}}(f)h_0^2$. To achieve it, we parameterize the energy spectrum using a smooth broken power-law and perform a Monte Carlo Markov Chain (MCMC) simulation with the released datasets. This approach enables us to quantitatively analyze the IGW energy spectra and investigate the related physics in the early universe such as inflation models and PBH formation. 

\section{The SGWB from scalar-induced gravitational waves}\label{sec:2}

We review the basic formalism for the IGWs \cite{Maggiore:1900zz, Boyle:2005se, Baumann:2007zm, Ananda:2006af} in this section. Second-order tensor perturbations are produced instantaneously when scalar modes re-enter the Hubble radius. In contrast to the linear-order perturbations, second-order perturbation theory introduces couplings among various $k$-modes of scalar, vector, and tensor perturbations in Fourier space. However, it is important to note that there is no direct mixing between the second-order scalar, vector, and tensor modes. In the context of second-order tensor perturbations, a noteworthy contribution arises from the quadratic dependence on linear-order scalar perturbations. This dependence gives rise to the generation of IGWs.
In perturbation theory, one can express the line element of perturbed spacetime in Newtonian gauge as \cite{Ananda:2006af, 1611.06130, 1804.08577}:
\begin{equation}
    \d s^2 = a^2 \Big[ -(1-2\Phi)\d\tau^2 + \big[ (1+2\Phi)\delta_{ij}+\frac{1}{2}h_{ij} \big] \d x^i\d x^j \Big]~, 
    \label{eq:lineelement}
\end{equation}
where $\tau$ is the conformal time, $\Phi$ is the Bardeen potential. $h_{ij}$ represents the second-order transverse and traceless tensor perturbation which is sourced by the non-linear couplings with linear-order scalar perturbations. The equation of motion for the Fourier component of second-order tensor perturbation $h_\mathbf{k} (\tau)$ is given by:
\begin{equation}
    h_{\bf k}^{\lambda \prime\prime}(\tau)+2\mathcal H h_{\bf k}^{\lambda \prime}(\tau)+k^{2}h_{\bf k}^{\lambda}(\tau)=S_{\bf k}^{\lambda}(\tau)~, 
    \label{eq:GWEoM}
\end{equation}
where ``$ ^\prime$'' denotes the derivatives with respect to the conformal time $\tau$, the comoving Hubble parameter $\mathcal{H}\equiv a'/a$, and $\lambda=+,\times$ represents the two polarizations of GWs. After horizon re-entry, the subsequent evolution of the tensor perturbations is scale dependent and determined by the time evolution of scalar modes. Especially, during the radiation dominated era (RD), the source term $S_\mathbf{k}(\tau)$ is expressed as:
\begin{equation}
    \begin{aligned}
        S_{\mathbf{k}}^\lambda(\tau) =&4 \int \frac{\mathrm{d}^3 \mathbf{p}}{(2 \pi)^3} \mathbf{e}^\lambda(\mathbf{k}, \mathbf{p})\left[3 \Phi_{\mathbf{p}} \Phi_{\mathbf{k}-\mathbf{p}}+\mathcal{H}^{-2} \Phi_{\mathbf{p}}' \Phi_{\mathbf{k}-\mathbf{p}}^{\prime}\right.\\
        &\left.+\mathcal{H}^{-1} \Phi_{\mathbf{p}}^{\prime} \Phi_{\mathbf{k}-\mathbf{p}}+\mathcal{H}^{-1} \Phi_{\mathbf{p}} \Phi_{\mathbf{k}-\mathbf{p}}^{\prime}\right],
        \label{eq:RDS}
    \end{aligned}
\end{equation}
where the projection is defined as $\mathbf{e}^\lambda(\mathbf{k},\mathbf{p})\equiv e^\lambda_{lm}(\mathbf{k})p_l p_m$ \cite{0912.5317}. Therefore, the correlator of IGWs is naturally ``induced'' by the correlator of linear scalar source:
\begin{equation}
    \begin{aligned}
\left\langle h_{\mathbf{k}}^\lambda(\tau) h_{\mathbf{k}^{\prime}}^s(\tau)\right\rangle= & \frac{1}{a^2(\tau)} \int^\tau \mathrm{d} \tau_1 \int^\tau \mathrm{d} \tau_2 g_k\left(\tau, \tau_1\right) g_k\left(\tau, \tau_2\right) \\
& \times a\left(\tau_1\right) a\left(\tau_2\right)\left\langle S_{\mathbf{k}}^\lambda\left(\tau_1\right) S_{\mathbf{k}^{\prime}}^s\left(\tau_2\right)\right\rangle~,
\label{eq:Correlator}
\end{aligned}
\end{equation}
where $g_k(\tau, \tau_1)$ denotes the Green's function which satisfies the equation of motion Eq.\eqref{eq:GWEoM}. According to the source term Eq.\eqref{eq:RDS}, the two-point correlator $\left\langle S_{\mathbf{k}}^\lambda S_{\mathbf{k}^{\prime}}^s\right\rangle$ can be expressed by the four-point correlator of curvature perturbations $\left\langle\mathcal{R}_\mathbf{p} \mathcal{R}_\mathbf{k-p} \mathcal{R}_\mathbf{q} \mathcal{R}_\mathbf{k^{\prime}-q}\right\rangle$ (or equivalently, the Bardeen potentials $\Phi_\mathbf{p}$). 

Accordingly, by performing the 2-point correlation of $h_\mathbf{k} (\tau)$ which is proportional to the correlator of scalar source $S_{\mathbf{k}}(\tau)$, one can arrive at the power spectrum of IGWs during RD: 
\begin{equation}
    \begin{aligned}
        \mathcal{P}_{h}^{\mathrm{RD}}(k,\tau)=&\int_{0}^{\infty}\mathrm{d}y\int_{|1-y|}^{1+y}\mathrm{d}x\left[{\frac{4y^{2}-(1+y^{2}-x^{2})^{2}}{4x y}}\right]^{2}\\
        &\times \mathcal{P}_{\mathcal{R}}(k x)\mathcal{P}_{\mathcal{R}}(k y)F_{\mathrm{RD}}(k\tau,x,y)~. 
    \end{aligned}
    \label{eq:RDPowerSpectrum}
\end{equation}
where the Green's functions in correlation Eq.\eqref{eq:Correlator} and the transfer function from curvature perturbation $\mathcal{R}$ to Bardeen potential $\Phi$ in RD are absorbed into $F_\mathrm{RD} (k\tau, x, y)$ \cite{1804.07732, 1804.08577, 1810.12224}, expressed as 
\begin{equation}
\begin{aligned}
    &F_{\rm RD}(k\tau,x,y)\\ 
    &=\frac{4}{81}\frac{1}{k^2\tau^2}\big[\cos^2(k\tau)\mathcal{I}_c^2+\sin^2(k\tau)\mathcal{I}_s^2+\sin(2k\tau)\mathcal{I}_c\mathcal{I}_s\big]~,
\end{aligned}
\end{equation}where the exact form of functions $\mathcal{I}_c(x,y)$ and $\mathcal{I}_s(x,y)$ can be seen, for example, in Refs.~\cite{1810.12224,Cai:2019jah}. 
Meanwhile, the dimensionless variables $x$ and $y$ are introduced as the ratio of momentum $x=|\mathbf{k}-\mathbf{p}| / k$ and $y=p / k$. The formalism Eq.\eqref{eq:RDPowerSpectrum} is the general expression to calculate the power spectrum of the GWs induced by the primordial curvature perturbation $\mathcal{R}$ when the associated Fourier modes of $\mathcal{R}$ re-enter the Hubble radius during RD. Despite the second-order tensor perturbations are suppressed in cases with a scale-invariant scalar power spectrum, it is still possible to observe GWs induced by the enhanced scalar perturbations which could trigger a formation of PBH after horizon re-entering. This kind of possible IGWs is what we are concerned about in this work.

When IGWs associated with PBH formation are triggered by the density perturbations, they become an SGWB which can be characterized by their energy spectrum $\Omega_\mathrm{GW}$. It is defined as the energy density of the GWs per unit logarithmic frequency. {During RD, it can be written as \cite{1804.07732, 1804.08577, 1810.12224}}:
\begin{equation}
    \Omega_\mathrm{GW}(\tau,k)=\frac{1}{24} \left(\frac{k}{\mathcal{H}}\right)^2\overline{\mathcal{P}^\mathrm{RD}_h(\tau,k)}~, 
\end{equation}
where the overline denotes the time average over several periods of the GWs. 
Taking into account the thermal history of the universe, the current observable energy spectrum of IGWs which is produced in RD is given by \cite{Zhao:2006mm,Pi:2020otn}
\begin{equation}
\textbf{}\begin{aligned}
\Omega_\mathrm{GW}^\mathrm{RD}(\tau_0,f)h_0^2=&1.6\times10^{-5}\left(\frac{g_{*s}(\tau_k)}{106.75}\right)^{-1/3}\\
&\times\left(\frac{\Omega_{r,0}h_0^2}{4.1\times 10^{-5}}\right)\Omega^\mathrm{RD}_\mathrm{GW}(\tau_\mathrm{eq},f)~, 
\label{eq:energyspectrumRD}
\end{aligned}
\end{equation}
where $\tau_\mathrm{eq}$ and $\tau_k$ denote the time at radiation-matter equality and the time of horizon crossing of the mode $k$, respectively, and $\Omega_{r,0}=5.38\times 10^{-5}$ is the current radiation energy density fraction \cite{1807.06209}. Here the comoving wavenumber $k$ is replaced by the physical frequency $f$ via $f=k/(2\pi a_0)\equiv k/(2\pi)$. 

Focusing on observations, the shapes of the SGWB energy spectra are essential. 
Different models or parameterizations of the power spectrum lead to distinct scale-dependent features in the energy spectrum of IGWs. Typically, these features can be effectively described by power-law relationships, with varying spectral indices $n_{\rm GW}(f)\equiv\d\ln{\Omega_{\rm GW}(f)}/\d\ln f$. For more universally exploring power spectra, we introduce a broken power-law parameterization to describe the energy spectrum as a well approximation~\cite{Cai:2019jah, Vaskonen:2020lbd}:
\begin{equation}\label{parameterization of GWs}
\Omega_{\text{GW}}(f) h_0^2 = A \frac{\alpha+\beta}{\beta\left(f / f_c\right)^{-\alpha}+\alpha\left(f / f_c\right)^{\beta}}~,
\end{equation}
where $A$ represents the amplitude of the energy spectrum at the critical frequency $f_c$. While $\alpha$ and $\beta$ are the indices of the spectrum, which describe the growth (blue-tilted) and decay (red-tilted) of the energy spectrum.

\section{Bayesian inference and results}\label{sec:3}

With the latest results of NANOGrav~\cite{NANOGrav:2023gor}, PPTA~\cite{Reardon:2023gzh}, and EPTA~\cite{Antoniadis:2023ott}, we apply the Bayesian inference method to determine the best fit of broken power-law parameterized spectrum in Eq.\eqref{parameterization of GWs}. 
We estimate the associated kernel density $\mathcal{L}_i$ for each frequency $f_i$ and its posteriors $\Omega_{\rm GW}(f_i)$. The likelihood is given by:
\begin{equation}
\mathcal{L}(\Theta)=\prod_{i=1} \mathcal{L}_{i}\left(\Omega_{\mathrm{GW}}\left(f_{i}; \Theta\right)\right),
\end{equation}
where the broken power-law energy spectrum at frequency $f_i$ with parameters $\Theta(A, f_c, \alpha, \beta)$ is represented by $\Omega_{\mathrm{GW}}(f_i; \Theta)$.
We adopt the MCMC sampler \texttt{emcee}~\cite{Foreman-Mackey:2012any} to sample the posterior probability. Priors of the model parameters are given in TABLE.~\ref{tab:best fit}.  

\begin{figure}[htbp]
\begin{center}
\includegraphics[width=0.98\linewidth]{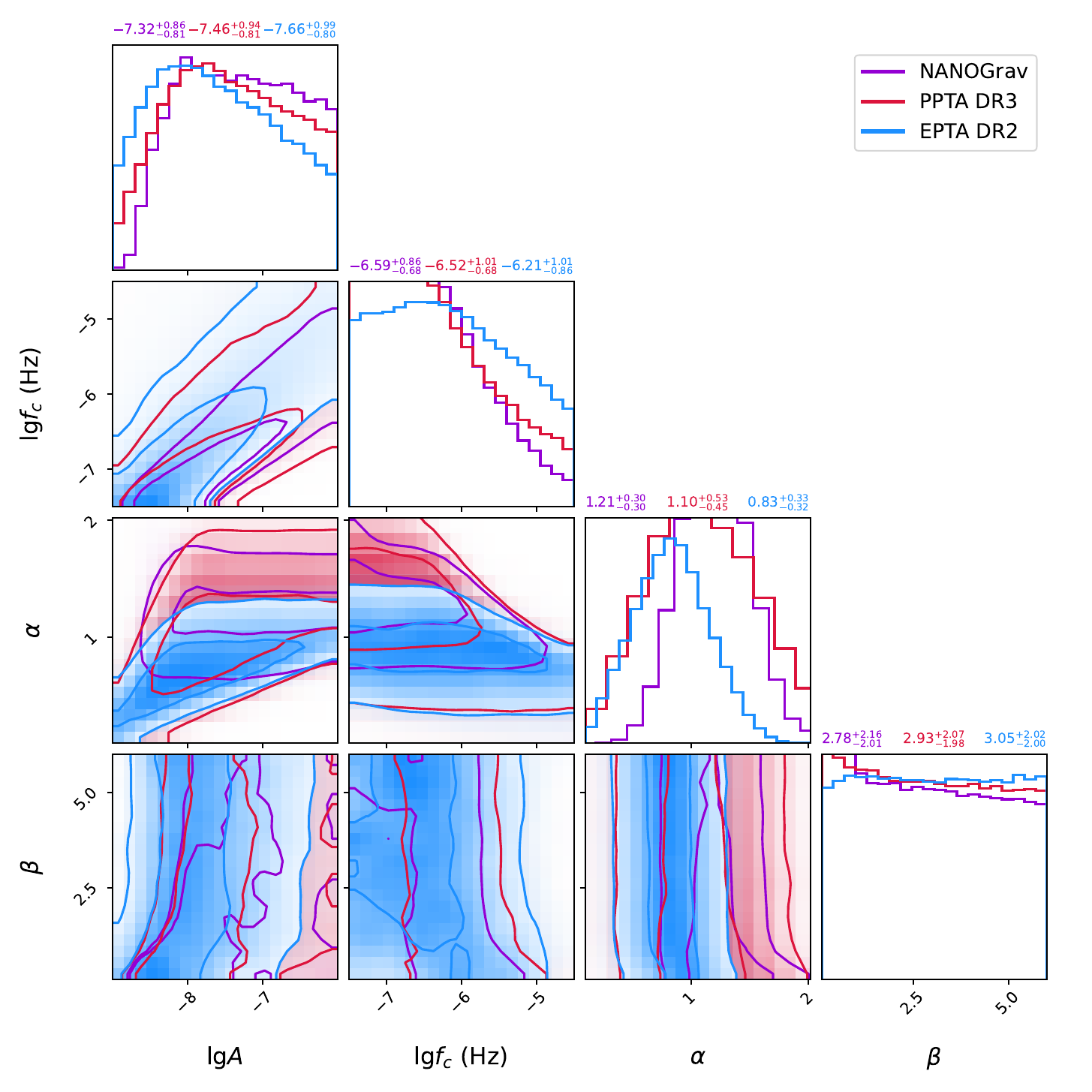}
\end{center}
\caption{Posterior distributions of the model parameters $({\rm lg}A, {\rm lg}f_c, \alpha, \beta)$ with $1\sigma$ and $2\sigma$ confidence level (C.L.). The distributions for NANOGrav, PPTA, and EPTA are represented by the purple, red, and blue regions respectively, and the best-fit results are listed in TABLE.~\ref{tab:best fit}. }
\label{fig_corner}
\end{figure}


\begin{table*}[!t]
\centering
\setlength{\tabcolsep}{.5em}
\caption{Prior distributions and parameter fitting results for NANOGrav, PPTA, and EPTA datasets at 95\% C.L. The priors for the indices $\alpha, \beta$ follow uniform distributions within the range $(0.1, 6.0)$, while the priors of the amplitude $A$ and critical frequency $f_c$ follow log-uniform distributions within $(-9.0,-6.0)$ and $(-7.5,-4.5)$, respectively. 
}
\begin{tabular}{ccccc}
\hline
  & lg$A$ &  lg$f_c$ & $\alpha$ & $\beta$ \\
\hline
Prior &  $\mathcal{U}(-9.0, -6.0)$ &  $\mathcal{U}(-7.5, -4.5)$ & $\mathcal{U}(0.1, 6.0)$ & $\mathcal{U}(0.1, 6.0)$ \\
NANOGrav &  $-7.32_{-0.81}^{+0.86}$ &  $-6.59_{-0.68}^{+0.86}$ & $1.21_{-0.30}^{+0.30}$ & $2.78_{-2.01}^{+2.16}$\\
PPTA &  $-7.46_{-0.81}^{+0.94}$ &  $-6.52_{-0.68}^{+1.01}$ & $1.10_{-0.45}^{+0.53}$ & $2.93_{-1.98}^{+2.07}$\\
EPTA &  $-7.66_{-0.80}^{+0.99}$ &  $-6.21_{-0.86}^{+1.01}$ & $0.83_{-0.32}^{+0.33}$ & $3.05_{-2.00}^{+2.02}$\\ 
\hline
\end{tabular}
\label{tab:best fit}
\end{table*}


\begin{figure}[htbp]
\begin{center}
\includegraphics[width=1\linewidth]{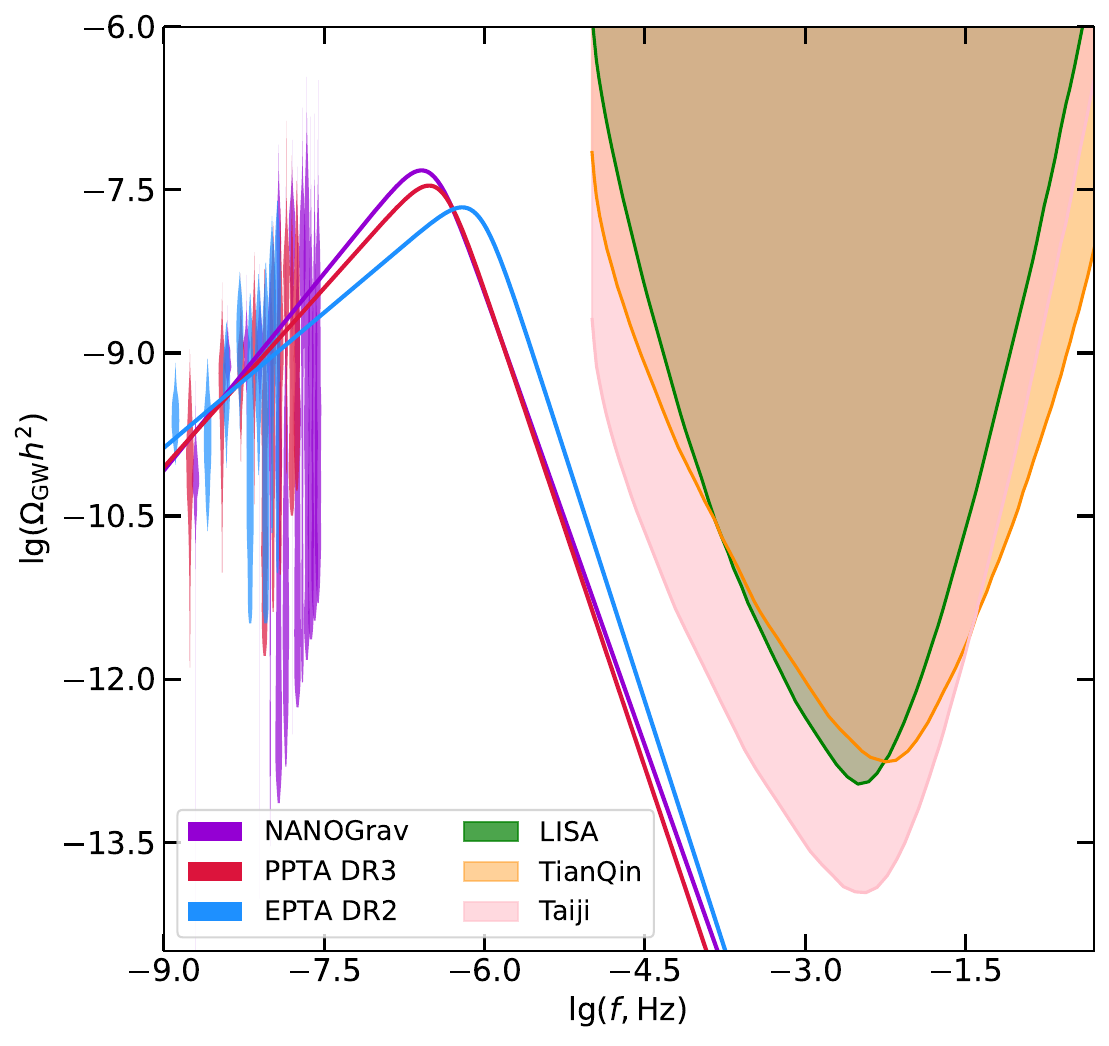}
\end{center}
\caption{The energy spectrum of the SGWB. The observations of NANOGrav, PPTA and EPTA are depicted in purple, red, and blue violin plots, respectively. The curves with the corresponding color are the IGW energy spectra represented by the best-fit results in FIG.~\ref{fig_corner}. 
The expected sensitivity curves of LISA~\cite{LISA:2017pwj}, TianQin~\cite{TianQin:2015yph} and Taiji~\cite{Ruan:2018tsw} are displayed in green, orange and pink shaded regions, respectively. These upcoming space-based GW observations are expected to test the ultra-violet (UV) tail of the IGW energy spectrum.}
\label{fig_Omega_h2}
\end{figure}

In FIG.~\ref{fig_corner}, we present the best-fit posterior distributions of parameters ($A$, $f_c$, $\alpha$, $\beta$) within 1$\sigma$ and 2$\sigma$ confidence regions, utilizing data from NANOGrav~\cite{NANOGrav:2023gor}, PPTA ~\cite{Reardon:2023gzh} and EPTA~\cite{Antoniadis:2023ott}. 
The uncertainties in datasets, coupled with the absence of UV observations, contribute to the suboptimal convergence of the parameters, particularly for $\beta$. Additionally, we provide the prior distributions and the 95\% C.L. best fits for parameters with NANOGrav, PPTA, and EPTA datasets in TABLE.~\ref{tab:best fit}. Moreover, FIG.~\ref{fig_Omega_h2} offers a comparison between the energy spectrum derived from the best-fit parameters and the corresponding observations. In the infrared (IR) region, $f<10^{-7}$Hz, NANOGrav and PPTA observe a steeper blue-tilted spectrum with index $\alpha>1.0$, while the spectrum observed by EPTA with $\alpha<1.0$.

It is important to note that typical parameterizations of scalar power spectra, such as log-normal and broken power-law peaks, which can arise from various small-scale enhanced inflation models~\cite{Kawasaki:1997ju,Kawasaki:2012wr,Cai:2018tuh,Chen:2019zza,Chen:2020uhe,Sasaki:2018dmp,Byrnes:2018txb,Mahbub:2019uhl,Ozsoy:2019lyy,Carrilho:2019oqg,Atal:2021jyo}, demonstrate similar behavior in the IR region. Specifically, the spectral index $n_{\rm GW}(f)$ is approximately described by $n_{\rm GW}=3-{2}/{\ln(f_c/f)}$. As $f\to 0$, corresponding to the IR limit, we encounter the well-known IR tail attributed to causality, where $\Omega_{\rm GW}$ scales as $f^3$. In the near-IR region, the narrow peak case predicts $n_{\rm GW}=2-2\ln(f_c/f)$~\cite{1910.09099}, while $\Omega_{\rm GW}$ follows the behavior of $\mathcal{P}_\mathcal{R}^2$ when the peak is broad~\cite{1910.09099,Pi:2020otn,2306.17149}. From our best-fit results, $f_c\gtrsim10^{-7}$Hz at 2$\sigma$, indicating that the observed PTA signal predominantly lies on the low-frequency side of the peak in the IGW energy spectrum, with $f/f_c\lesssim\mathcal{O}(0.1)$. In this case, $\alpha$ roughly equals to $n_{\rm GW}$. More specifically, the signal resides in the near-IR region ($\alpha \lesssim 2$) according to the best-fit $\alpha$ (see TABLE.~\ref{tab:best fit}). 

Our results indicate limitations in extracting information about the source of SGWB within the PTA frequency band, primarily due to the signal predominantly residing in the IR region. To gain a comprehensive understanding of the origin of SGWB, it becomes crucial to explore features both around the peak and on the UV side of the GW energy spectrum. When the SGWB energy spectrum has a relatively gentle UV tail, corresponding to smaller values of $\beta$ in our model, there is potential for future space GW observation projects, such as LISA~\cite{LISA:2017pwj}, TianQin~\cite{TianQin:2015yph}, and Taiji~\cite{Ruan:2018tsw}, to detect the associated signals. In FIG.~\ref{fig_Omega_h2}, we emphasize that a potential detection of this UV tail by LISA/TianQin/Taiji could carry significant implications. Specifically, it might lead to the disfavoring of models proposing a narrow log-normal peak power spectrum, which predicts a substantially large $\beta$.

\section{Summary and Discussion}\label{sec:4}

The collaborative efforts of various PTA experiments, including NANOGrav, PPTA, EPTA, and CPTA, have collectively presented compelling evidence of SGWB within the nano-Hertz frequency band~\cite{NANOGrav:2023gor,Reardon:2023gzh,Antoniadis:2023ott,Xu:2023wog}. Such background offers promising prospects for investigating cosmological sources originating in the early universe. In particular, IGWs emerge as a potential source contributing to the SGWB. IGWs are typically associated with amplified primordial curvature perturbations that could lead to the production of PBHs.  As a cosmological-originated SGWB, IGWs provide valuable insights into the fundamental physics in the early universe and the origin of PBHs. 
 
The currently released data strongly implicates a blue-tilted spectrum within the PTA frequency band, specifically in the range of $[10^{-9}, 10^{-7}]$ Hz. It is crucial to reiterate that understanding the shape of GW energy spectra is paramount for uncovering the origins of IGWs in the primordial universe. Consequently, in this study, we introduce a broken power-law parameterization of GW energy spectrum and employ Bayesian inference to analyze datasets released by NANOGrav, PPTA, and EPTA. The results from the MCMC simulation, depicted in FIG.~\ref{fig_Omega_h2} and TABLE.~\ref{tab:best fit}, indicate that the maximum of $\Omega_\mathrm{GW} h_0^2$ will likely exceed $10^{-8}$, with the corresponding peak frequency $f_c$ surpassing $10^{-8}$Hz. If these signals indeed originate from IGWs, they would provide significant insights into the peak position of the primordial power spectrum of scalar perturbations and the associated mass of the resultant PBHs. It is noteworthy that the broken power-law parameterization of the GW spectrum is not only applicable to the scalar-induced SGWB but can also be extended to other cosmological and astrophysical sources. For the purposes of this letter, our analysis focuses solely on the IGW source, leaving the exploration of other sources for future research.  

Given the inverse-square relation between the frequency of IGWs and the mass of related PBHs \cite{Cai:2019jah}, our results predict the quasi-monochromatic mass of PBHs should be smaller than $\mathcal{O}(10^{-1}) M_\odot$. In this mass range, assuming Gaussian curvature perturbations would lead to an issue of PBH overproduction~\cite{Franciolini:2023pbf}. Generally, two strategies can be employed to alleviate this concern: incorporating non-Gaussian perturbations \cite{Cai:2018dig,Liu:2023ymk, Franciolini:2023pbf, Wang:2023ost, Atal:2021jyo} and introducing non-trivial inflation \cite{Harigaya:2023pmw}. Introducing a negative non-Gaussian parameter $F_{\text{NL}}$ helps mitigate this issue, but a positive $F_{\text{NL}}$ 
exacerbates the overproduction of PBHs~\cite{Franciolini:2023pbf,Wang:2023ost,Liu:2023ymk}. When $F_{\text{NL}} < 0$, the influence of non-Gaussianities on the shape of the spectrum is not significant \cite{Cai:2018dig}. In contrast, $f_{\text{PBH}}$ is extremely sensitive to non-Gaussianities. Consequently, while a negative $F_{\text{NL}}$ alleviates the overproduction issue, our MCMC results regarding the shape of the energy spectrum still serve as a reasonably good approximation.

Furthermore, in frequency bands beyond the peak of the spectrum, IGWs anticipate a red-tilted tail. This UV tail is expected to undergo testing via forthcoming space-based GW observations working within $f \gtrsim 10^{-6}~{\rm Hz}$ frequency band. 
Thus, in the near future, IGWs could be verified by the joint observations of space-based laser interferometer experiments and PTA. 
The nature of the energy spectrum has the potential to significantly enhance our comprehension of small-scale physics in the early universe, providing valuable insights into the likelihood of the existence of PBHs.  \\

\section*{Acknowledgements} 
We thank Qing-Guo Huang, Shi Pi, Misao Sasaki, Zhao-Qiang Shen, and Chi Zhang for the valuable discussion. 
This work is supported in part by the National Key R\&D Program of China (2021YFC2203100), by CAS Young Interdisciplinary Innovation Team (JCTD-2022-20), by NSFC (11875113, 11961131007, 12261131497, 12003029, 11833005, 12192224), by Innovation and Talent Introduction Base for Ministry of Education and Ministry of Science and Technology for "Observational and Theoretical Research on Dark Matter and Dark Energy" (B23042), by Fundamental Research Funds for Central Universities, by the Disposizione del Presidente INFN n. 24433 in INFN Sezione di Milano, by China Postdoctoral Science Foundation under grant No. 2023TQ0355, by CSC Innovation Talent Funds, by USTC Fellowship for International Cooperation, by USTC Research Funds of the Double First-Class Initiative, by CAS project for young scientists in basic research (YSBR-006).

\end{document}